\documentclass{qgbws-procs9x6}

\def\al{\alpha}
\def\be{\beta}
\def\ga{\gamma}

\def\ep{\epsilon}

\def\et{\eta}

\def\ka{\kappa}
\def\la{\lambda}

\def\rh{\rho}

\def\Ph{\Phi}

\def\cl{{\mathcal L}}
\def\fr#1#2{{{#1} \over {#2}}}
\def\prt{\partial}

\def\frac#1#2{{\textstyle{{#1}\over {#2}}}}

\def\lsim{\mathrel{\rlap{\lower4pt\hbox{\hskip1pt$\sim$}}
    \raise1pt\hbox{$<$}}}
\def\gsim{\mathrel{\rlap{\lower4pt\hbox{\hskip1pt$\sim$}}
    \raise1pt\hbox{$>$}}}

\def\curl#1{\vec\nabla\times\vec #1}

\def\grad#1{\vec\nabla #1}
\def\etal {{\it et al.}}

\newcommand{\beq}{\begin{equation}}
\newcommand{\eeq}{\end{equation}}
\newcommand{\bea}{\begin{eqnarray}}
\newcommand{\eea}{\end{eqnarray}}
\newcommand{\rf}[1]{(\ref{#1})}

\def\kaf{k_{AF}}
\def\kf{k_{F}}

\def\kfi{(k_{F})_{\ka\la\mu\nu}}
\def\kt{\tilde k}
\def\ko{\tilde \ka_{o+}}
\def\ke{\tilde \ka_{e-}}
\def\ktr{\tilde \ka_{\rm tr}}

\begin{document}

\title{Lorentz-Violating Electromagnetostatics}

\author{Quentin G.\ Bailey}

\address{Physics Department, \\
Indiana University, \\ 
Bloomington, IN 47405, USA\\ 
E-mail: qbailey@indiana.edu}

\maketitle

\abstracts{
In this talk, the stationary limit
of Lorentz-violating electrodynamics
is discussed.
As illustrated by some simple examples,
the general solution includes
unconventional mixing of electrostatic and magnetostatic effects.  
I discuss a high-sensitivity null-type measurement,
exploiting Lorentz-violating electromagnetostatic
effects, that could improve existing limits
on parity-odd coefficients for Lorentz violation
in the photon sector.}

\section{Introduction}
\label{Introduction}

Experiments to date have shown that Lorentz symmetry
is an exact symmetry of all known forces
in nature.  However, many ongoing experiments
are searching for small violations of Lorentz symmetry
that could arise in the low-energy limit of a unified theory
of nature at the Planck scale.\cite{cpt01}  
Much of the analysis of these experiments 
is performed within a theoretical framework 
called the Standard-Model Extension (SME).\cite{ck} 
The SME is an effective field theory
that extends the Standard Model (SM) and general relativity
to include small violations of particle Lorentz and CPT symmetry
while preserving observer Lorentz symmetry and the
coordinate invariance of physics.
The CPT and Lorentz-violating terms
in the SME lagrangian
have coupling coefficients with Lorentz indices
which control the Lorentz violation,
and can be viewed as low-energy remnants
of the underlying physics at the Planck scale.\cite{kps}
Tests of this theory include ones with
photons,\cite{cavexpt1,cavexpt2,kmphot,photonexpt,photonexpt2}
electrons,\cite{eexpt}
protons and neutrons,\cite{ccexpt}
mesons,\cite{hadronexpt}
muons,\cite{muexpt}
neutrinos,\cite{nu}
and the Higgs.\cite{higgs}

In the photon sector of the minimal SME,
recent Lorentz symmetry tests have focused
on the properties of electromagnetic waves
in resonant cavities and propagating {\it in vacuo}.
I show in this talk,
however,
that there are unconventional effects associated with
the stationary,
non-propagating limit of the photon sector.
I also discuss experimental possibilities
based on these effects in high-sensitivity null-type measurements.
A detailed discussion of this topic 
is contained in Ref. \refcite{ems}.\footnote{See Ref. \refcite{photonth}
for theoretical literature on the photon sector of the SME.}  

\section{Framework}
\label{Framework}

The lagrangian density for the photon sector of the minimal SME
can be written as
\bea
\cl &=& -\fr 1 4 F_{\mu\nu}F^{\mu\nu}
- \frac 1 4 \kfi F^{\ka\la}F^{\mu\nu}
\nonumber\\ & &
+\frac 1 2 (\kaf)^\ka\ep_{\ka\la\mu\nu}A^\la F^{\mu\nu}
- j^\mu A_\mu.
\label{L_em1}
\eea
In this equation,
$j^\mu = (\rh, \vec J)$ is the 4-vector current source
that couples to the electromagnetic 4-potential $A_\mu$,
and $F_{\mu\nu} \equiv \prt_\mu A_\nu - \prt_\nu A_\mu$
is the electromagnetic field strength.
From this definition the conventional homogeneous Maxwell
equations are automatically satisfied.
The coefficients $\kfi$ and $(\kaf)^\ka$ are assumed constant
and control the CPT and Lorentz violation.
The current $j_\mu$ is taken to be conventional,
thus assuming Lorentz violation is only present in the photon sector.
The CPT-odd coefficients $(\kaf)^{\ka}$
are stringently bounded by cosmological observations
and are set to zero in this analysis.\cite{photonexpt}  
The lagrangian \rf{L_em1}
yields the inhomogeneous equations of motion
\beq
\prt_\al{F_\mu}^\al
+(\kf)_{\mu\al\be\ga}\prt^\al F^{\be\ga}
+j_{\mu}=0.
\label{eqmotphot1}
\eeq 
These equations can be written as Maxwell equations
in terms of $\vec D, \vec H, \vec E$ and $\vec B$ by defining
appropriate vacuum constituency relations.\cite{kmphot,ems}
Equation \rf{eqmotphot1} can be generalized
to include regions of isotropic matter.
The usual linear response
of matter to applied fields is modified
by Lorentz violation
and additional matter coefficients appear
in the constituency relations.\cite{ems}  

\section{Electromagnetostatics}
\label{Electromagnetostatics}

The stationary solutions of the
modifed Maxwell equations \rf{eqmotphot1}
{\it in vacuo} will 
satisfy the time-independent equation of motion
\beq
\kt^{j\mu k\nu} \prt_j \prt_k
A_{\nu}(\vec x) = j^{\mu}(\vec x),
\label{eqmotphot3}
\eeq
where the coefficients $\kt^{j\mu k\nu}$ are defined by
\beq
\kt^{j\mu k\nu} =
\et^{jk}\et^{\mu\nu}-\et^{\mu k} \et^{\nu j}+2(\kf)^{j\mu k\nu}.
\label{diffop}
\eeq
From the homogeneous Maxwell equations
the electrostatic and magnetostatic fields
can be written in terms of
the 4-potential $A^\mu = (\Ph, A^j)$ as
$\vec E = - \grad \Ph$ and $\vec B = \curl A$.
The metric terms in Eq. \rf{diffop} are the conventional terms 
that split \rf{eqmotphot3} into separate equations
for the scalar potential from charge density and
the vector potential from current density.  
The presence of the $(\kf)^{j\mu k\nu}$
term implies that a static charge density
generates a small vector potential
and a modified scalar potential
and similarly a steady-state current density generates
a small scalar potential and a modified vector potential.
Electrostatics and magnetostatics,
while distinct in the conventional case,
become convoluted in the presence of Lorentz violation.
Discussing the static limit of Lorentz-violating electrodynamics
therefore requires the simultaneous
treatment of both electric and magnetic phenomena.\cite{fn1}

To obtain a general solution for the potentials $\Ph$ and $\vec A$
I introduce Green functions $G_{\mu\al}(\vec x,\vec x')$
that solve Eq.\ \rf{eqmotphot3} for a point source.  
Once a suitable Green theorem that
incorporates the differential operator in Eq.\ \rf{eqmotphot3}  
is found, the formal solution 
can be constructed for a spatial region $V$
in terms of the Green functions,
the 4-current density and the values
of the potential on the boundary $S$.
The general solution is
\bea
A_{\la}(\vec x) &=&
\int_V d^3x' G_{\mu\la}(\vec x',\vec x) j^{\mu}(\vec x')  
\nonumber\\ 
& - & \int_S d^2S' \hat n'^j [G_{\mu\la}(\vec x',\vec x) 
\kt^{j\mu k\nu}\prt'_k A_{\nu}(\vec x')
\nonumber\\ & &
\qquad \qquad
- A_{\mu}(\vec x') \kt^{j\mu k\nu}\prt'_k G_{\nu\la} (\vec x',\vec x)].
\label{potsol}
\eea
Manipulation of Eq. \rf{potsol}
reveals four classes of boundary conditions
that establish unique solutions for
the electric and magnetic fields:
$(\Ph, \hat n \times \vec A)$,
$(\Ph, \hat n \times \vec H)$,
$(\hat n \cdot \vec D ,
\hat n \times \vec A)$,
$(\hat n \cdot \vec D,
\hat n \times \vec H )$.
With each of these sets
of boundary conditions there are corresponding constraints
on the Green functions.\cite{ems}  
The mixing of $\Ph$ and $\vec A$ in the boundary conditions
comes from the unconventional definitions
of the fields $\vec D$ and $\vec H$.\cite{kmphot,ems}
The solution \rf{potsol} can be
generalized to regions of isotropic matter
using a modified version of Eq. \rf{diffop}.\cite{ems}

\section{Applications}  
\label{Applications}

As a first application of Eq. \rf{potsol}
I consider the case of boundary conditions at infinity
in which the surface terms are dropped.
Imposing the Coulomb gauge,
the explicit form for the Green functions can be extracted
from fourier decomposition in momentum space.\cite{ems}
For the case of a point charge
at rest at the origin the scalar potential\cite{kmphot} 
and vector potential are given by
\bea
\Ph (\vec x) &=&  \fr q {4\pi|\vec x|}
\Big(
1 - (\kf)^{0j0k} \hat x^j \hat x^k
\Big) ,
\nonumber\\
A^j(\vec x) &=& \fr q {4\pi |\vec x|}
\Big(
(\kf)^{0kjk} - (\kf)^{jk0l} \hat x^k \hat x^l
\Big).
\label{b}
\eea
Equation \rf{b} shows explicitly that a point charge at rest produces
a magnetic field in the presence of Lorentz violation, which
is obtained from $B^j = \ep^{jkl}\prt^k A^l$.\cite{ems}

Consider now an example
motivated by a possible experimental application.  
I seek the fields from a magnetic source surrounded
by a conducting shell.
In the idealized solution presented
the magnetic source is a sphere of radius $a$
and uniform magnetization $\vec M$ surrounded
by a grounded conducting shell of radius $R>a$.
The fields for this configuration
can be obtained from \rf{potsol}
using the $(\Ph, \hat n \times \vec A)$ set
of boundary conditions and treating the
magnetic source as a current density $\vec J = \curl M$.
The leading order solution for the scalar potential $\Ph$
in the region $a<r<R$,
where $r$ is the radial coordinate from the center of the sphere,
is given by
\bea
\Ph (\vec x) = {\hat r \cdot \ko
\cdot \vec m \over 4\pi }
\left( {1 \over r^2}
- {r \over R^3} \right),
\label{finsoln}
\eea
where $\vec m =  4\pi a^3\vec M/3$.
Here we have made use of the zero-birefringence
approximation that $(\ko)^{jk} = (\kf)^{0jpq}\ep^{kpq}$
is an anti-symmetric matrix.\cite{kmphot,ems}
The solution \rf{finsoln} becomes modified
in the more realistic scenario
with the magnet consisting of matter obeying
Lorentz-violating matter constituency relations.\cite{ems}
 
\section{Experiment}
\label{Experiment}

Recent experiments in the photon sector
are least sensitive to $\ko$ and $\ktr = -\fr 2 3 (\kf)^{0j0j}$.
This is due to the parity-odd nature of the
corresponding Lorentz-violating effects from $\ko$
and the scalar nature of $\ktr$ to which
recent experiments are only indirectly sensitive.
The setup of the second example in Sec. \ref{Applications}
is designed to be directly sensitive to parity-odd effects.
It can be seen directly from \rf{finsoln}
that the scalar potential, if taken to be the observable,
is proportional to $\ko$.\cite{wolf}
A suitable experiment would 
measure the potential from Eq. \rf{finsoln}
in the space between the magnet 
and outer shell ($a<r<R$).
The outer conducting shell then serves to sheild the apparatus from
external electric fields.
For an estimate of the sensitivity that
might be attainable I assume the source
is a ferromagnet with strength $10^{-1}$ T near its surface
and the voltage sensitivity is at the level of nV.
A null measurement could then achieve
a sensitivity $\ko \lsim 10^{-15}$.
This represents an improvement by $10^4$
over the best existing sensitivities.\cite{cavexpt2}

Equation \rf{finsoln} is written in the
laboratory frame.  
Since this frame is fixed to the earth
it is not inertial on the time scale
of the earth's rotation and revolution.
The resultant time dependence of the signal 
can be obtained by transforming
the laboratory-frame coefficients $(\ko)^{jk}_{\rm lab}$
to a Sun-centered inertial frame following Ref.\ \refcite{kmphot}.
Thus, with upper-case letters denoting Sun-centered coordinates,
\bea
(\ko)_{\rm lab}^{jk}
&=&T_0^{jkJK}(\ko)^{JK}
+2T_1^{kjJJ} \ktr
\nonumber\\ &&
+(T_1^{kjJK} -T_1^{jkJK}) (\ke)^{JK}
\label{timedep} ,
\eea
where $T_0^{jkJK}=R^{jJ}R^{kK}$
and $T_1^{jkJK}= R^{jP}R^{kJ} \ep^{KPQ} \be^Q$
are tensors containing the time dependence
from the rotations $R^{jJ}$
and the boost $\be^J$.
One can also consider
rotating the entire apparatus
to produce a signal with a shorter time variation,
which may increase sensitivity and reduce systematics.
With these considerations
one can attain time-dependent sensitivity
to all three independent components of $\ko$
and time-dependent sensitivity to $\ktr$
suppressed by a single power of $|\vec \be| \simeq 10^{-4}$.


\begin{thebibliography}{xx}

\bibitem{cpt01}
For summaries of recent Lorentz tests,
see, for example,
V.A.\ Kosteleck\'y, ed.,
{\it CPT and Lorentz Symmetry II},
World Scientific, Singapore, 2002.

\bibitem{ck}
D.\ Colladay and V.A.\ Kosteleck\'y,
Phys.\ Rev.\ D {\bf 55}, 6760 (1997);
Phys.\ Rev.\ D {\bf 58}, 116002 (1998);
V.A.\ Kosteleck\'y,
Phys.\ Rev.\ D {\bf 69}, 105009 (2004);
R.\ Bluhm and V.A.\ Kosteleck\'y,
Phys.\ Rev.\ D {\bf 71}, 065008 (2005).

\bibitem{kps}
V.A.\ Kosteleck\'y and S.\ Samuel,
Phys.\ Rev.\ D {\bf 39}, 683 (1989);
V.A.\ Kosteleck\'y and R.\ Potting,
Nucl.\ Phys.\ B {\bf 359}, 545 (1991).

\bibitem{cavexpt1}
J.\ Lipa \etal,
Phys.\ Rev.\ Lett.\ {\bf 90}, 060403 (2003).

\bibitem{cavexpt2}
H.\ M\"uller \etal,
Phys.\ Rev.\ Lett.\ {\bf 91}, 020401 (2003);
P.\ Wolf \etal,
Gen.\ Rel.\ Grav.\ {\bf 36}, 2351 (2004);
Phys.\ Rev.\ D {\bf 70}, 051902 (2004).

\bibitem{kmphot}
V.A.\ Kosteleck\'y and M.\ Mewes,
Phys.\ Rev.\ D {\bf 66}, 056005 (2002).

\bibitem{photonexpt}
S.M.\ Carroll \etal,
Phys. Rev. D {\bf 41}, 1231 (1990);
M.P.\ Haugan and T.F.\ Kauffmann,
Phys. Rev. D {\bf 52}, 3168 (1995).

\bibitem{photonexpt2}
V.A.\ Kosteleck\'y and M.\ Mewes,
Phys.\ Rev.\ Lett.\ {\bf 87}, 251304 (2001).

\bibitem{eexpt}
H.\ Dehmelt \etal,
Phys.\ Rev.\ Lett.\ {\bf 83}, 4694 (1999);
R.\ Mittleman \etal,
Phys.\ Rev.\ Lett.\ {\bf 83}, 2116 (1999);
G.\ Gabrielse \etal,
Phys.\ Rev.\ Lett.\ {\bf 82}, 3198 (1999);
R.\ Bluhm \etal,
Phys.\ Rev.\ Lett.\ {\bf 82}, 2254 (1999);
Phys.\ Rev.\ Lett.\ {\bf 79}, 1432 (1997);
Phys.\ Rev.\ D {\bf 57}, 3932 (1998);
D.\ Colladay and V.A.\ Kosteleck\'y,
Phys.\ Lett.\ B {\bf 511}, 209 (2001);
B.\ Heckel,
in Ref.\ \refcite{cpt01};
L.-S.\ Hou \etal,
Phys.\ Rev.\ Lett.\ {\bf 90}, 201101 (2003);
R.\ Bluhm and V.A.\ Kosteleck\'y,
Phys.\ Rev.\ Lett.\ {\bf 84}, 1381 (2000);
H.\ M\"uller \etal,
Phys. Rev. D {\bf 68}, 116006 (2003);
B.\ Altschul, 
Phys. Rev. D {\bf 70}, 056005 (2004).

\bibitem{ccexpt}
L.R.\ Hunter \etal,
in V.A.\ Kosteleck\'y, ed.,
{\it CPT and Lorentz Symmetry}, 
World Scientific, Singapore, 1999;
D.\ Bear \etal,
Phys.\ Rev.\ Lett.\ {\bf 85}, 5038 (2000);
D.F.\ Phillips \etal,
Phys.\ Rev.\ D {\bf 63}, 111101 (2001);
M.A.\ Humphrey \etal,
Phys.\ Rev.\ A {\bf 68}, 063807 (2003);
Phys.\ Rev.\ A {\bf 62}, 063405 (2000);
F.\ Can\`e \etal,
Phys.\ Rev.\ Lett.\ {\bf 93} 230801 (2004);
V.A.\ Kosteleck\'y and C.D.\ Lane,
Phys.\ Rev.\ D {\bf 60}, 116010 (1999);
J.\ Math.\ Phys.\ {\bf 40}, 6245 (1999);
R.\ Bluhm \etal,
Phys.\ Rev.\ Lett.\ {\bf 88}, 090801 (2002);
Phys.\ Rev.\ D {\bf 68}, 125008 (2003).

\bibitem{hadronexpt}
KTeV Collaboration,
in Ref.\ \refcite{cpt01};
OPAL Collaboration,
Z.\ Phys.\ C {\bf 76}, 401 (1997);
DELPHI Collaboration,
preprint DELPHI 97-98 CONF 80 (1997);
BELLE Collaboration,
Phys.\ Rev.\ Lett.\ {\bf 86}, 3228 (2001);
BaBar Collaboration,
Phys.\ Rev.\ Lett.\ {\bf 92}, 142002 (2004);
FOCUS Collaboration,
Phys.\ Lett.\ B {\bf 556}, 7 (2003);
V.A.\ Kosteleck\'y and R.\ Potting,
Phys.\ Rev.\ D {\bf 51}, 3923 (1995);
V.A.\ Kosteleck\'y,
Phys.\ Rev.\ Lett.\ {\bf 80}, 1818 (1998);
Phys.\ Rev.\ D {\bf 61}, 016002 (2000);
Phys.\ Rev.\ D {\bf 64}, 076001 (2001).

\bibitem{muexpt}
V.W.\ Hughes \etal,
Phys.\ Rev.\ Lett.\ {\bf 87}, 111804 (2001);
R.\ Bluhm \etal,
Phys.\ Rev.\ Lett.\ {\bf 84}, 1098 (2000).

\bibitem{nu}
S.\ Coleman and S.L.\ Glashow,
Phys.\ Rev.\ D {\bf 59}, 116008 (1999);
V.\ Barger \etal,
Phys.\ Rev.\ Lett.\ {\bf 85}, 5055 (2000);
J.N.\ Bahcall \etal,
Phys.\ Lett.\ B {\bf 534}, 114 (2002);
I.\ Mocioiu and M.\ Pospelov,
Phys.\ Lett.\ B {\bf 537}, 114 (2002);
A.\ de Gouv\^ea, Phys.\ Rev.\ D {\bf 66}, 076005 (2002);
G.\ Lambiase,
Phys.\ Lett.\ B {\bf 560}, 1 (2003);
V.A.\ Kosteleck\'y and M.\ Mewes,
Phys.\ Rev.\ D {\bf 69}, 016005 (2004);
Phys.\ Rev.\ D {\bf 70}, 031902 (2004);
Phys.\ Rev.\ D {\bf 70}, 076002 (2004);
S.\ Choubey and S.F.\ King,
Phys.\ Lett.\ B {\bf 586}, 353 (2004);
A.\ Datta \etal,
Phys.\ Lett.\ B {\bf 597}, 356 (2004).

\bibitem{higgs}
D.L.\ Anderson \etal,
Phys.\ Rev.\ D {\bf 70}, 016001 (2004);
E.O.\ Iltan,
Mod.\ Phys.\ Lett.\ A {\bf 19}, 327 (2004).

\bibitem{photonth}
R.\ Jackiw and V.A.\ Kosteleck\'y,
Phys.\ Rev.\ Lett.\ {\bf 82}, 3572 (1999);
M. P\'rez-Victoria, JHEP {\bf 0104}, 032 (2001);
V.A.\ Kosteleck\'y \etal,
Phys.\ Rev.\ D {\bf 65}, 056006 (2002);
C.\ Adam and F.R.\ Klinkhamer,
Nucl.\ Phys.\ B {\bf 657}, 214 (2003);
V.A.\ Kosteleck\'y \etal,
Phys.\ Rev.\ D {\bf 68}, 123511 (2003);
H.\ M\"uller \etal,
Phys. Rev. D {\bf 67}, 056006 (2003);
T.\ Jacobson \etal,
Phys.\ Rev.\ D {\bf 67}, 124011 (2003);
V.A.\ Kosteleck\'y and A.G.M.\ Pickering,
Phys.\ Rev.\ Lett.\ {\bf 91}, 031801 (2003);
R.\ Lehnert,
Phys.\ Rev.\ D {\bf 68}, 085003 (2003);
G.M.\ Shore,
Contemp.\ Phys.\ {\bf 44}, 503 {2003};
B.\ Altschul, 
Phys.\ Rev.\ D {\bf 69}, 125009 (2004);
Phys.\ Rev.\ D {\bf 70}, 101701 (2004);
Nucl.\ Phys.\ B {\bf 705}, 593 (2005);
hep-th/0402036; 
R.\ Lehnert and R.\ Potting,
Phys.\ Rev.\ Lett.\ {\bf 93}, 110402 (2004);
Phys.\ Rev.\ D {\bf 70}, 125010 (2004);
R.\ Lehnert,
J.\ Math.\ Phys.\ {\bf 45}, 3399 (2004).

\bibitem{ems}
Q.G.\ Bailey and V.A.\ Kosteleck\'y,
Phys.\ Rev.\ D {\bf 70}, 076006 (2004).

\bibitem{fn1}
See also C.\ L\"ammerzahl, CPT04 proceedings.
 
\bibitem{wolf}
For an alternate method see P. Wolf \etal,
Phys.\ Rev.\ D {\bf 71}, 025004 (2005); 
M. Tobar, CPT04 proceedings.

\end{thebibliography}
\end{document}